\documentclass{article}

\AtBeginDocument{%
  \providecommand\BibTeX{{%
    \normalfont B\kern-0.5em{\scshape i\kern-0.25em b}\kern-0.8em\TeX}}}

\usepackage{caption}
\usepackage{subcaption}
\usepackage{todonotes}
\usepackage{pgf-umlsd}
\usepackage{tikz}
\usetikzlibrary{arrows.meta}
\usepackage{amsthm}
\newcommand*{\defeq}{\stackrel{\text{def}}{=}}

\usepackage{xcolor}

\usepackage{amsmath}
\usepackage{cleveref}

\usepackage{url}

\definecolor{myblue}{HTML}{0000FF} % ETH
\definecolor{myred}{HTML}{FF0000} % wstETHCRV
\definecolor{mypurple}{HTML}{7F00FF} % wstETHCRV-gauge
\definecolor{myorange}{HTML}{CC6600} % dForce: USX Token
\definecolor{mydarkbrown}{HTML}{663300} % wstETH
\definecolor{mygreen}{HTML}{00994D} % dForce wstETHCRV-gauge tokens

\theoremstyle{definition}
\newtheorem{definition}{Definition}[section]

\usepackage{float}
\usepackage{placeins}

\begin{document}

\title{Instrumenting Transaction Trace Properties in Smart Contracts: Extending the EVM for Real-Time Security}

\author{Zhiyang Chen\\
\texttt{jeff@zircuit.com}
\and
Jan Gorzny\\
\texttt{jan@zircuit.com}
\and 
Martin Derka\\
\texttt{martin@zircuit.com}
}

%\author{Lars Th{\o}rv{\"a}ld}
%\affiliation{%
%  \institution{The Th{\o}rv{\"a}ld Group}
%  \streetaddress{1 Th{\o}rv{\"a}ld Circle}
%\email{larst@affiliation.org}

%\renewcommand{\shortauthors}{Chen et al.}


\date{15 July 2024}

\maketitle

\begin{abstract}
  In the realm of smart contract security, transaction malice detection has been able to leverage properties of transaction traces to identify hacks with high accuracy.
  However, these methods cannot be applied in real-time to revert malicious transactions.
  Instead, smart contracts are often instrumented with some safety properties to enhance their security.
  However, these instrumentable safety properties are limited and fail to block certain types of hacks such as those which exploit read-only re-entrancy. This limitation primarily stems from the Ethereum Virtual Machine's (EVM) inability to allow a smart contract to read transaction traces in real-time. Additionally, these instrumentable safety properties can be gas-intensive, rendering them impractical for on-the-fly validation. To address these challenges, we propose modifications to both the EVM and Ethereum clients, enabling smart contracts to validate these transaction trace properties in real-time without affecting traditional EVM execution. We also use past-time linear temporal logic (PLTL) to formalize transaction trace properties, showcasing that most existing detection metrics can be expressed using PLTL. We also discuss the potential implications of our proposed modifications, emphasizing their capacity to significantly enhance smart contract security.

\end{abstract}

\section{Introduction}\label{sec:introduction}

Smart contracts are self-executing programs that run on blockchain networks. 
They are essential infrastructure for blockchain based applications like Decentralized Financial (DeFi) \cite{DBLP:conf/sp/ZhouXECWWQWSG23} applications. 
As of July 1, 2024, the Total Value Locked (TVL) in 3,871 DeFi protocols has reached an impressive \$$83.57$ billion USD~\cite{defillama2}. 
However, the landscape faces significant challenges. Security hacks are a major concern for the integrity of smart contracts. Malicious hackers can exploit various vulnerabilities by executing hack transactions, potentially resulting in the theft of millions of dollars. As of July 1, 2024, financial losses due to security attacks on DeFi protocols have surpassed \$$8.3$ billion USD~\cite{defillama}.

Current methods to prevent these attacks on the fly are limited. The prevailing strategy involves instrumenting safety properties within smart contracts to automatically revert transactions if these properties are violated. Yet, this approach has failed to completely prevent real-world exploits. 
Conversely, various detection metrics have been developed to analyze transaction execution traces and identify malicious activities. These metrics typically examine certain properties of transaction trace and may also reference external sources like price oracles to assess potential threats. These detection metrics have been proved effective in identifying malicious transactions. For example, Forta~\cite{forta} attack detectors have detected 75\% of major on-chain hacks in 2023~\cite{fortadetect}. Despite their analytical capabilities, these detection metrics cannot be used on the fly and are not capable of stopping attacks in real-time.

The lack of real-time prevention is primarily due to the limitations of the Ethereum Virtual Machine (EVM) and the absence of a comprehensive system architecture that can validate trace properties on the fly. The EVM restricts users' ability to define complex smart contract safety properties and instrument them in smart contracts as a defense mechanism. These limitations are inherent to Ethereum's design and exist in many EVM-compatible blockchains. 
We propose a novel industrial solution to allow the definition of complex trace properties and validate them in real-time to prevent or suspend malicious transactions. Our solution fills the gap between current effective detection techniques and the low efficiency of real-time prevention.

Unlike traditional software systems, where the execution trace could be private and inaccessible to the public, the execution traces (transactions) of smart contracts is completely transparent and accessible to everyone. These traces are accessible by any node in the network even when the transactions are pending and have not been included in a block. Other works such as front-running~\cite{eskandari2020sok} utilizes the transparency of the transaction trace to detect profitable transactions and front-run them. However, front-running may not always be possible (e.g., if the block builder is the one who introduced the malicious transaction in private).
Our work leverages the transparency of the transaction trace to define and validate trace properties in real-time.

\noindent 
\textbf{Contributions} This paper makes the following contributions:
\begin{itemize}
    \item  We propose an innovative industrial solution that modifies the EVM to support the definition and real-time validation of trace properties, aiming to halt or suspend malicious transactions on the fly.
    \item We introduce a formalism of trace properties using past-time linear temporal logic, demonstrating that it can effectively represent the majority of trace properties utilized in current real-world detection metrics.
    \item We provide comprehensive discussions on practical use cases, potential impacts, and prospective directions for future research derived from our findings.
\end{itemize}

\section{Related Work}\label{sec:related}
In this section, we review the existing literature and prior developments 
in the field of blockchain security and malice detection, 
identifying the gaps that our research addresses.

\subsection{Smart Contract Safety Property Runtime Validation}

There is a significant body of work exploring safety properties that can be directly instrumented in smart contracts to revert malicious transactions~\cite{chen2024demystifying, zhou2020ever, liu2022learning, liu2022finding, callens2024temporarily}. Although these approaches represent best practices in both academia and industry for stopping hacks in real-time, they face practical challenges when applied by developers to smart contracts. Specifically, the runtime guards utilized in these works can be gas-intensive, thereby increasing the cost for users to interact with these instrumented contracts. Our work allows developers to define trace properties that can be validated in real-time without the need for runtime guards, thereby reducing the gas cost for users.

\subsection{Transaction Trace Properties for Malice Detection}

Detecting anomalous transactions on blockchain platforms, such as Ethereum, has been a central focus of research aimed at improving the security and integrity of smart contracts. 
Prior works like \textit{TxSpector}~\cite{zhang2020txspector} pioneered the bytecode-level analysis of Ethereum transactions to identify attacks. This was followed by advancements like The Eye of Horus~\cite{ferreira2021eye} and Time-Travel Investigation~\cite{wu2022time}, which further refined attack detection mechanisms for smart contracts and blockchain transactions. The detection metrics used in these works can be translated into the trace properties defined in our work to not only detect but also prevent malicious transactions in real-time. In industry, Forta Network~\cite{forta} attack detectors have detected 75\% of major on-chain hacks in 2023~\cite{fortadetect}.

\subsection{Linear Temporal Logic for Smart Contract Formal Specification}

Linear Temporal Logic (LTL) is extensively employed to define the temporal properties of smart contracts. It characterizes the safety and liveness of transition-system models, which are validated using model checkers~\cite{boxin2024research, alqahtani2020formal, atzei2019developing, bai2018formal, bartoletti2019verifying, molina2018implementation}. However, existing research focuses on verifying the correctness of smart contracts rather than identifying malicious transactions as they occur. Our research shifts this focus towards transaction trace properties. Specifically, we examine properties associated with the execution trace up to a certain step (defined as ``hooks''), reasoning about a determined path rather than all possible paths.

% \todo[inline]{Incorporate this attack as motivating example. Say that the ``read-only'' part of it was reentered in a way that the smart contract only detect by trivially marking everything as read-only.}

\section{Motivating Example}~\label{sec:example}
\begin{figure}[bt]
    \centering
    \includegraphics[width=0.75\textwidth]{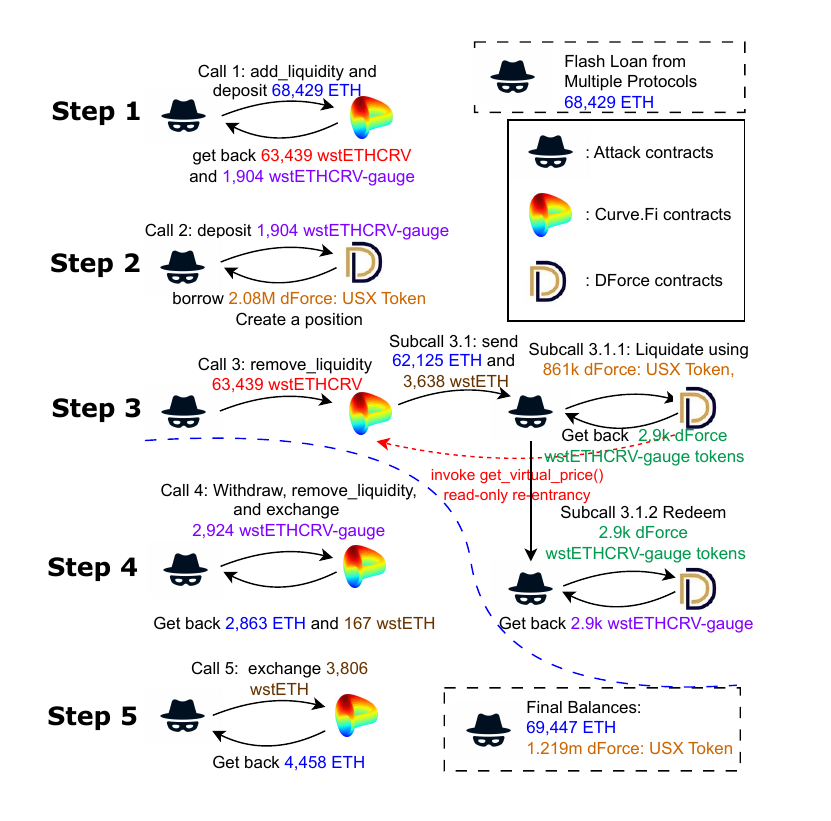}
    \caption{The dForce incident, illustrated. The entire
    exploit takes place during a single transaction on Arbitrum. The view function \texttt{get\_virtual\_price()} is accessed when executing the \texttt{remove\_liquidity()} function, and is used as an oracle by dForce.}
    \label{fig:dforce}
  \end{figure}

The dForce attack on February 13, 2023 exploited a read-only reentrancy vulnerability within the dForce's integration with Curve Finance on the Arbitrum and Optimism blockchains, leading to a substantial financial impact of approximately \$3.6 million USD~\cite{dforce}. 
Figure~\ref{fig:dforce} details the sequence of events in the exploit on Arbitrum, as pieced together from various post-mortem analyses~\cite{dforce, dforce-pm} and on-chain data~\cite{dforce-tx}.

\noindent
\textbf{Step 0.} The attacker begins by borrowing \textcolor{myblue}{68,429} \textcolor{myblue}{ETH} (worth approximately \$105 Million USD) 
via flash loans across multiple flash loan providers. 

\noindent
\textbf{Step 1.} The attacker deposits the borrowed \textcolor{myblue}{68,429} \textcolor{myblue}{ETH} into the Curve Finance wstETH/ETH pool through the \texttt{add\_liquidity()} function and receives \textcolor{myred}{63,439} \textcolor{myred}{wstETHCRV} and \textcolor{mypurple}{1,904} \textcolor{mypurple}{wstETHCRV-gauge} tokens in return. 

\noindent
\textbf{Step 2.} Using the newly acquired \textcolor{mypurple}{wstETHCRV-gauge} tokens, the attacker creates leveraged positions within dForce, borrowing \textcolor{myorange}{2.08M dForce: USX Token}. This step is crucial as it establishes a big debt position that can be liquidated later in the exploit.

\noindent
\textbf{Step 3.} The attacker invokes the \texttt{remove\_liquidity()} function from the Curve Finance contract, burning their \textcolor{myred}{63,439}  \textcolor{myred}{wstETHCRV} tokens to withdraw \textcolor{myblue}{ETH} and \textcolor{mydarkbrown}{wstETH}. This action triggers the fallback function within their malicious contract when \textcolor{myblue}{ETH} is sent to the attacker. It is important to note that the \texttt{get\_virtual\_price()} function, which computes asset prices, is a read-only function and uses the total token supply for its calculation. However, \texttt{remove\_liquidity()} function does not strictly follow the Checks-Effects-Interactions pattern, and the \texttt{get\_virtual\_price()} function is called before the total supply changes are made. Thus, during this re-entry, the return value of \texttt{get\_virtual\_price()} is wrong and significantly smaller than the actual price.
This incorrect pricing, subsequently accessed by dForce as an oracle, enables the attacker to liquidate positions at a falsely low cost. Using just \textcolor{myorange}{2.08M dForce: USX Token}, the attacker liquidates not only their own position in \textbf{Step 2} but also liquidates other users' positions, getting a liquidation reward of \textcolor{mygreen}{2.9k dForce wstETHCRV-gauge tokens}. The hacker then exchanges these tokens for \textcolor{mypurple}{2.9k wstETHCRV-gauge tokens}. 

\noindent
\textbf{Step 4.} The \textcolor{mypurple}{dForce wstETHCRV-gauge tokens} are then exchanged back to \textcolor{myblue}{ETH} and \textcolor{mydarkbrown}{wstETH} through Curve.

\noindent
\textbf{Step 5.} The attacker completes the exploit by converting all \textcolor{mydarkbrown}{wstETH} to \textcolor{myblue}{ETH}, including the liquidated assets, and repaying the initial flash loans. The final profit, after all transactions, totals 1018 \textcolor{myblue}{ETH} (worth approximately \$1.57 Million USD) and  \textcolor{myorange}{1.219m dForce: USX Token} (worth approximately \$1.22 Million USD).

From the analysis of the exploit steps, it becomes apparent that the underlying cause of the incident was the misplaced trust by dForce in the oracle price provided by Curve Finance's \texttt{get\allowbreak\_virtual\_price()} function. The attacker strategically timed their re-entry into the Curve contract during the liquidity removal process, allowing them to manipulate this price. This artificial inflation of the oracle price enabled the attacker to liquidate positions within dForce at significantly distorted rates, exacerbating the financial impact of the attack.

\subsection{Challenges for Runtime Validation}

To prevent re-entrancy and oracle manipulation, smart contracts typically employ invariant guards such as re-entrancy guards~\cite{ozreentrancy} and oracle deviation checks (e.g., \cite{chen2024demystifying}). 
In the dForce incident, re-entrancy guards would not have been effective since it was Curve Finance's contracts that were re-entered, not dForce's. dForce could not detect re-entrancy in Curve Finance unless Curve updated its implementation and added re-entrancy guards in their contracts. Oracle deviation checks could have been partially effective; dForce would have noticed a significant deviation from previous oracle prices when checked against the manipulated prices. However, attackers could still access and manipulate the oracle price to lesser extents, repeating their attack vectors multiple times, albeit with reduced profits and increased transaction costs. Nonetheless, dForce could still suffer significant exploitation. Furthermore, oracle deviation checks require continuous monitoring by developers to ensure that the recorded oracle prices remain current. Infrequent checks could lead to outdated data and numerous transaction reversions.

Due to the limitations of the EVM, a contract cannot access the execution traces of other contracts executed before it. For instance, dForce contracts could not determine whether the Curve Finance contract, which served as their oracle, had been re-entered. Moreover, they cannot access, via the EVM, that the transaction involved a flash loan of over \textcolor{myblue}{68,429 ETH} from various providers, which should be considered highly abnormal and suspicious. 

Given these challenges, we propose a new system that extends the EVM, enabling smart contracts to access transaction trace prior to their execution. Our system permits developers to integrate "hooks" into smart contracts and specify trace properties that must be met at these points. For instance, dForce could implement a trace property to verify whether a transaction involves a flash loan or exhibits a re-entrancy pattern at a hook right before a token transfer. If a violation occurs, the transaction is immediately reverted. This mechanism could have preemptively thwarted and mitigated incidents like the dForce hack.

% % adding hooks 
% This flexibility would enable developers to design custom invariant guards. For example, dForce could implement a guard to verify if any accessed contract had been re-entered or if the transaction involved a flash loan. We believe that with this innovation, incidents like the dForce hack could be effectively mitigated.

\section{Example Safety Properties}\label{sec:safety-properties}

There are safety properties that can be instrumented directly in smart contracts to revert malicious transactions, as detailed in Section~\ref{sec:related}.
%Despite these advances, as noted in the previous section, these EVM-instrumentable safety properties fall short in preventing hacks, such as read-only re-entrancy attacks, primarily due to the limitations inherent in the EVM. %%% jgorzny: too long of a sentence
Despite these advances, as noted in the previous section, these EVM-instrumentable safety properties fall short in preventing hacks.
Attacks such as read-only re-entrancy attacks are difficult to prevent primarily due to the limitations inherent in the EVM.

A variety of approaches for detecting transaction malice exist both in academia~\cite{zhang2020txspector} and in industry~\cite{hypernative, forta}. These methods effectively identify transaction trace properties; however, they typically report malicious transactions post-execution, by which point financial losses have already occurred. In Section~\ref{sec:design}, we discuss a novel system architecture that transcends EVM constraints, allowing for the pre-execution enforcement of transaction trace properties by integrating hooks into the EVM.

To illustrate the practical implications and motivate our proposed design, we conducted a systematic analysis of $188$ malice detectors from the Forta network~\cite{forta}. 
We highlight $3$ transaction trace properties that, while currently utilized for malice detection, are impossible or challenging to instrument within the EVM. Our discussion centers on how these properties can be leveraged in our system to thwart attacks like the dForce incident.

\noindent
\textbf{Flashloan Detection~\cite{flashloanDetector}:}
This transaction trace property monitors for flashloan usage during critical protocol functions, such as minting or redeeming. Under standard EVM architecture, a victim contract cannot read the entire call stack to detect if it is being triggered via a flashloan callback function, hence unable to proactively flag such transactions as malicious. Our enhanced system, however, allows victim contracts to read all prior execution steps including the entire call stack through embedded hooks upon invocation. This capability enables the system to automatically detect and flag the involvement of flashloan providers and the extent of flashloan used, thus preventing malicious transactions in real-time.

\noindent
\textbf{Re-entrancy Detection~\cite{reentrancyDetector}:}
This transaction trace property is designed to identify re-entrancy during the execution of token transfer functions. It checks whether a contract is re-entered within the call stack during a token transfer operation and flags the transaction if re-entrancy is detected. In our system, for instance, when dForce initiates the transfer of 2.9k wstETHCRV-gauge tokens to a hacker, integrated hooks empower the dForce contract to detect and flag any re-entrancy activities in real-time, including those targeting specific functions like the Curve Fi's \texttt{get\_virtual\_price()}.

\noindent
\textbf{TVL Abrupt Change Detection~\cite{tvlChangeDetector}:}
This transaction trace property actively monitors for significant changes in the TVL within protocols. 
In the EVM, detecting such changes requires reading token prices from oracle contracts where it could be potentially outdated or manipulable. Unlike the conventional system, our proposed architecture is not confined to smart contract-based oracles; it instead supports integrating reliable endpoints or third-party oracles. This enhancement allows for real-time, accurate monitoring of TVL shifts, enabling the system to promptly flag transactions that result in abrupt TVL changes as malicious.

\section{Logic}\label{sec:logic}

We use Past-time Linear Temporal Logic (PLTL) to specify properties of smart contracts. PLTL allows us to express temporal properties about the execution of smart contracts, including conditions that must hold at different points in time. Notably, PLTL has the same expressiveness as Linear Temporal Logic (LTL)~\cite{gabbay1980temporal}. Moreover, algorithms exist to translate PLTL formulas into LTL formulas, such as those presented in~\cite{gabbay1989declarative}. We choose PLTL here because, when reasoning about smart contract execution, it is often more intuitive to consider past-time properties (i.e., the outcomes of previous executions) when determining whether the current smart contract is under attack.

\begin{definition}[$PLTL$]
Let $AP$ be a set of atomic propositions, and let $q,p \in AP$. PLTL is defined with the followng syntax:
$$
\phi, \psi ::= \neg \phi \mid (\phi \land \psi) \mid \mathbf{X} \phi \mid (\phi \, \mathbf{U} \, \psi) \mid (\phi \, \mathbf{S} \, \psi) \mid \mathbf{X}^{-1} \phi \mid p \mid q \mid \ldots
$$
\end{definition}

The syntax of PLTL includes all the elements of LTL, with additional past-time modalities. The semantics of LTL fragment of PLTL are the classical ones. For the past-time modalities, given a path $\sigma$ and a position $i$, we have:

\begin{itemize}
    \item \textbf{since}:  $\sigma, i \models \phi \, \mathbf{S} \, \psi$ if and only if there exists $k \leq i$ such that $\sigma, k \models \psi$ and for all $j$ with $k < j \leq i$, $\sigma, j \models \phi$.
    \item \textbf{previously}: $\sigma, i \models \mathbf{X}^{-1} \phi$ if and only if $i \geq 1$ and $\sigma, i-1 \models \phi$.
\end{itemize}

The classical abbreviations $\mathbf{F}$ (\textbf{eventually}), $\mathbf{G}$ (\textbf{always}), and their past-time counterparts, $\mathbf{F}^{-1}$ (\textbf{Once}) and $\mathbf{G}^{-1}$ (\textbf{Historically}), can be defined in terms of the other operators: 
(1) $\mathbf{F} \varphi \defeq \top \mathbf{U} \varphi$
(2) $\mathbf{G} \varphi \defeq \neg \mathbf{F} \neg \varphi$
(3) $\mathbf{F}^{-1} \varphi \defeq \top \mathbf{S} \varphi$
(4) $\mathbf{G}^{-1} \varphi \defeq \neg \mathbf{F}^{-1} \neg \varphi$ 

\noindent
\textbf{Extend PLTL with Quantifiers}
To match the expressiveness of PLTL with the detection metrics used in practice, we extend PLTL with quantifiers. We introduce the universal quantifier $\forall$ and the existential quantifier $\exists$ to reason about all or some paths, respectively. The syntax of the extended PLTL is as follows:
$$
\phi, \psi ::= ... \mid \forall x \, \phi \mid \exists x \, \phi \mid \ldots
$$
where $x$ ranges over a domain of variables specific to the transaction trace (e.g., addresses or function selectors).

\subsection{Expressing Real-world Transaction Trace Properties with Past-Time Linear Temporal Logic}
We aim to explore how many real-world detection metrics can be expressed using PLTL.
Forta~\cite{forta}, a malice detection service provider, employs numerous detectors to identify anomalies and allows users to define custom detection rules. We systematically collected $188$ Forta attack detectors listed in ~\cite{fortadetect}. Among these, we found that $42$ can be used for runtime validation, meaning they can be instrumented in a smart contract to serve as invariant guards. However, it is important to note that while these can be instrumented, it does not necessarily mean they are practical as runtime guards due to potentially high gas costs, which users are reluctant to incur. This is why only a few runtime validation techniques are used in practice, and these are typically very simple invariant guards.

Furthermore, we discovered that $186$ detectors can be expressed using PLTL. These detectors are all related to reasoning about the past trace of the current transaction. There are $2$ detectors that cannot be expressed using PLTL. The first involves checking the Matic~\footnote{The Matic token is the native cryptocurrency of the Polygon network, used for paying transaction fees, participating in governance decisions, and securing the network through staking.} price, and the second involves checking pending transactions. Expressing these two detectors requires accessing external resources such as offchain oracles and the Ethereum client mempool. While incorporating this external information into our transaction trace properties could be an interesting topic, it is beyond the scope of this work.

Our study has demonstrated the substantial expressiveness of PLTL. In the following, we demonstrate how to use PLTL to define three sample transaction trace properties of smart contracts as shown in Section~\ref{sec:safety-properties}. We assume developers add hooks before every token transfer invocation in their smart contract code, denoted as $\mathit{TokenTransfer(x)}$ where $x$ is the amount of tokens. We also assume that the transaction trace are already parsed into an invocation tree, and the $\mathit{CallStack}$ represents a list of tuples of contract addresses and function selectors.

\noindent
\textbf{Flashloan Detection:} 
Flashloan detection involves ensuring that no flashloan functions are invoked before executing critical protocol functions such as $\mathit{TokenTransfer(x)}$. Let $\mathit{InFlashLoanProviders(c, s)}$ be a predicate that is true if the contract $c$ and selector $s$ represent a function that provides flashloan.
\begin{align*}
    \psi &\equiv  \mathit{TokenTransfer(x)} \rightarrow \\
    &  \neg \mathbf{F}^{-1} \left(  \forall(c ,s) \in \mathit{CallStack}.  \neg \mathit{InFlashLoanProviders}(c, s)  \right)
\end{align*}

\noindent
\textbf{Re-entrancy Detection:} 
Re-entrancy detection involves ensuring that no contract has been entered twice before during the execution of token transfer functions. 
\begin{align*}
    \psi &  \equiv \mathit{TokenTransfer(x)} \rightarrow  \\
    &  \neg \mathbf{F}^{-1} \left(  \forall(c_i ,s_i), (c_j, s_j), i \not = j \in \mathit{CallStack}. \neg (c_i = c_j \land s_i = s_j)  \right)
\end{align*}

\noindent
\textbf{TVL Abrupt Change Detection:} 
TVL abrupt change detection involves monitoring significant changes in the TVL within protocols. We define the TVL as the difference between the sum of past deposits and the sum of past withdrawals. Let 
$p$ be a threshold value that represents the maximum allowable change in TVL. The property can be defined as follows:
\begin{align*}
    \psi &\equiv \mathit{TokenTransfer(x)} \rightarrow \\
    & x < p \cdot( \text{sum(deposits)} - \text{sum(withdrawals)} )
\end{align*}

\begin{figure*}[ht]
\centering
\includegraphics[width=0.6\textwidth]{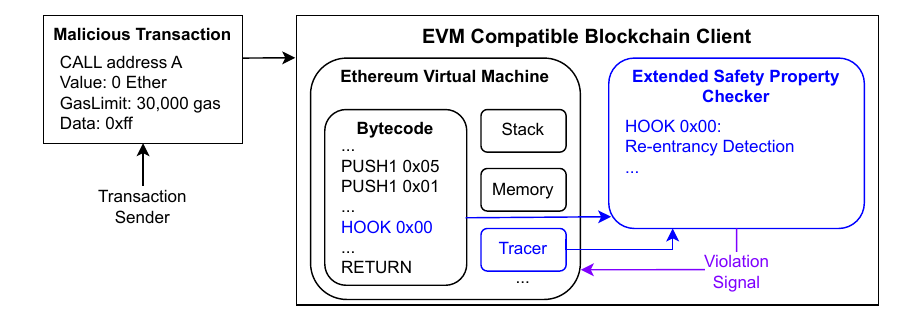}
\caption{Modified Geth client to support checking transaction trace properties. The client is extended with a new module that allows smart contracts to check transaction trace safety properties. The module is responsible for maintaining the state of the contract and checking the properties. The module is triggered by hooks in the EVM execution.}
\label{fig:geth}
\end{figure*}

\section{Extending EVM to Check Transaction Trace Properties}\label{sec:design}

In the previous sections, we formalized the use of PLTL for 
defining transaction trace properties and demonstrated how these properties can be 
utilized to enhance the security of smart contracts.
This section delves into the practical implementation of such a system within a blockchain environment, 
using the Geth execution client as an example.
As shown in Figure~\ref{fig:geth}, to enable the real-time verification of transaction trace properties, 
we propose extending the Geth client with a new module. 
This module is designed to maintain the state of the smart contract 
and check the defined safety properties during execution. 
The module is activated by hooks integrated into the EVM execution process, 
allowing it to monitor and enforce safety properties dynamically.

\subsection{EVM Execution Modifications}
A new module \textit{tracer} is added to the Geth client. It will track 
every execution steps of the EVM and collects the transaction trace. Note 
similar module has existed in the Geth client for debugging purposes. 
The tracer, when encountering a hook, will send all the collected transaction
trace to the hook for checking. 

The hook is defined by the smart contract developers and is embedded in the
contract code.
These hooks can be triggered at any step during the VM execution, enabling the real-time 
checking of safety properties. This implementation requires creating a modified 
fork of the EVM that supports these additional capabilities. Here we introduce a 
new opcode \texttt{HOOK} that allows users to add hooks to their contract code. 
When the EVM encounters this opcode, it triggers the safety property checking module, 
which then evaluates the defined properties. 

\subsection{Use Cases}
When writing smart contracts, developers can incorporate hooks into their code, 
potentially using varying keywords specific to each smart contract language. 
Alongside the contract, developers are required to author an additional code file dedicated 
to defining the transaction trace properties they wish to monitor in each hook. 
These trace properties are designed to only read from the transaction trace and are 
restricted from modifying the blockchain state directly. This feature serves 
as a runtime guard, enabling developers to preemptively block malicious transactions. 
Furthermore, it allows for the exploration of transaction trace properties that are beyond 
the current implementation capabilities of the EVM. This proactive approach enhances security 
and extends the functional breadth of smart contract monitoring.

This approach can be used in multiple ways.
First, such an opcode could be integrated into Ethereum nodes so that smart contracts compiled with it are not rejected.
This will require nodes to upgrade their software and accept this new method.
One could sidestep this by replacing the the \texttt{HOOK} opcode with a no-op that compiles but signals to nodes running such an implementation to perform additional analysis and only include any calling transactions if the analysis reports no issues.
For example, smart contract developers could call \texttt{keccak256(bytes(``HOOK''))} and relevant nodes could watch for the call to trigger analysis (though this particular example requires a relatively large amount of gas).
However, rollups and other layer two solutions \cite{DBLP:conf/fc/GudgeonMRMG20} provide a perfect opportunity to implement this system.
In particular, rollups which try to enforce security at the sequencer level can use a modified version of their execution client to support this additional opcode \cite{sls-arxiv}.

\subsection{Overhead Analysis}
The primary bottlenecks in blockchain systems are consensus and storage~\cite{li2020securing}. 
Notably, our approach does not introduce additional consensus requirements, as the safety properties 
are checked locally by the module within the EVM. Additionally, since checking safety properties does not 
modify the blockchain state, it does not exacerbate storage issues. Once a hook is triggered, 
we can spawn another process to check whether the safety properties are satisfied. 
This process can be run in parallel with the EVM execution, ensuring that the safety properties 
are verified without affecting the performance of the blockchain.

\section{Conclusion}\label{sec:conclusion}

In this paper, we point out an interesting industrial fact: 
while the detection techniques of smart contract hacks using transaction trace properties 
have been very successful,
these techniques are not able to be applied in real-time to prevent the hacks, mainly due
to the limitations of the EVM on limiting smart contracts to read
transaction trace on the fly. 
We formalize the transaction trace properties using 
the past-time linear temporal logic, and demonstrate most detection metrics can be expressed using it. 
We show that how to modify EVM and the Ethereum client 
to allow smart contracts to check these properties on the fly. 
We also provide insightful discussions 
on the implications and implementations of the proposed system.

% \begin{acks}
% TODO
% \end{acks}

%\newpage
\bibliographystyle{unsrt}
\bibliography{ref}

%%
% %% If your work has an appendix, this is the place to put it.
% \appendix
% \input{author-notes}

\end{document}